\documentclass[osajnl,twocolumn,showpacs,superscriptaddress,10pt]{revtex4-1}
\usepackage{amsmath,amssymb,graphicx,subeqnarray}
\begin{document}
\title{Frequency-Bin Entanglement with Tunable Phase}

\author{Xianxin Guo}
\affiliation{Department of Physics, The Hong Kong University of Science and Technology, Clear Water Bay, Kowloon, Hong Kong, China}

\author{Peng Chen}
\affiliation{Department of Physics, The Hong Kong University of Science and Technology, Clear Water Bay, Kowloon, Hong Kong, China}

\author{Chi Shu}
\affiliation{Department of Physics, The Hong Kong University of Science and Technology, Clear Water Bay, Kowloon, Hong Kong, China}

\author{M. M. T. Loy}
\affiliation{Department of Physics, The Hong Kong University of Science and Technology, Clear Water Bay, Kowloon, Hong Kong, China}

\author{Shengwang Du}\email{Corresponding author: dusw@ust.hk}
\affiliation{Department of Physics, The Hong Kong University of Science and Technology, Clear Water Bay, Kowloon, Hong Kong, China}

\begin{abstract}
We describe a technique to produce narrow-band photon pairs with frequency-bin entanglement, whose relative phase can be tuned using linear polarization optics. We show that, making use of the polarization-frequency coupling effect, the phase of a complex polarizer can be transferred into the frequency entanglement.
\end{abstract}


\maketitle 


Entanglement is an essential nature of compound quantum systems \cite{QuantumEntanglement_2009, PanRMP2012}. For wide-band photon pairs produced from spontaneous parametric down conversion (SPDC) in a nonlinear crystal, entanglement between discrete frequency bands can be manipulated with a dispersive prism pair and a spatial light modulator \cite{Bernhard_PRA2013, SilberbergPRL2005}. However this method can not be applied to narrow-band biphotons whose bandwidth and band gap are on the order of 10 MHz. Zeilinger \textit{et al}. showed that the polarization entanglement can be transferred into frequency using a polarizing beam splitter (PBS) and polarization projections \cite{Zeilinger_PRL2009}. In this configuration, the phase adjustment is done in the polarization entanglement before the PBS.

Here we describe a different approach to create narrow-band frequency entanglement with tunable phase. Following our recent demonstration of generating narrow-band biphotons with polarization-frequency-coupled hyperentanglement, we find that it is possible to transfer the polarization-induced phase into that of the frequency entanglement \cite{CShu_arxiv2014}. Different from the work by Zeilinger \textit{et al}. \cite{Zeilinger_PRL2009}, our scheme does not require an initial polarization entanglement from the source. We make use of the polarization-frequency coupling to transfer the phase of a complex polarizer into the frequency entanglement.

\begin{figure}[h]
\includegraphics[width=8.5cm]{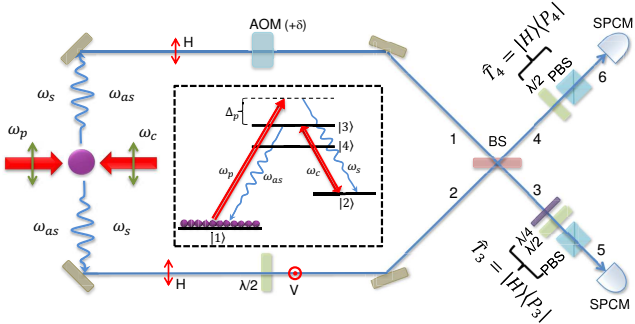}
\caption{\label{fig:system setup}(Color online). Schematic of experimental setup for producing narrow-band biphotons with frequency-bin entanglement. The phase-matched paired photons are generated from spontaneous four-wave mixing in laser-cooled $^{85}$Rb atoms with a right-angle pump-coupling geometry. The relevant $^{85}$Rb energy levels are $|1\rangle=|5S_{1/2},F=2\rangle$, $|2\rangle=|5S_{1/2},F=3\rangle$, $|3\rangle=|5P_{3/2},F=3\rangle$, and $|4\rangle=|5P_{3/2},F=2\rangle$.}
\end{figure}

The schematic of our experimental setup and related atomic energy level  diagram are depicted in Fig.~\ref{fig:system setup}. We produce narrow-band biphotons from spontaneous four-wave mixing (SFWM) in a laser-cooled $^{85}$Rb atomic ensemble with a right-angle geometry \cite{CLiu_PRA2012}. The coupling laser ($\omega_c$, 3 mW, diameter 2mm) is on resonance with the transition $|2\rangle\rightarrow|3\rangle$, and the pump laser ($\omega_p$, 40mW, diameter 2mm) is far blue detuned from the transition $|1\rangle\rightarrow|3\rangle$ by $\Delta_p=2\pi\times$3033 MHz. With the counter-propagating pump and coupling laser beams, phase-matched backward paired Stokes ($\omega_s$) and anti-Stokes ($\omega_{as}$) photons are spontaneously produced with exchange symmetry in paths 1 and 2: Stokes photons go to path 1 and anti-Stokes photons to path 2, and vice versa. To ensure there is no polarization entanglement pre-required in our configuration, we pass both ports (1 and 2) to horizontal (H) linear polarizers. Then we shift the optical angular frequencies of the photons in path 1 by $\delta=2\pi\times100$ MHz using an acousto-optical modulator (AOM), and adjust the photons in path 2 to be vertically (V) polarized using a half-wave plate (HWP). The photons from paths 1 and 2 are incident to a 50:50 beam splitter (BS). The photons from the two output ports 3 and 4 of the BS pass through two polarizers (P$_3$ and P$_4$) and are detected by two single-photon counting modules (SPCM) with a time-bin width of 1 ns. Further experimental details are described in \cite{CShu_arxiv2014}.

In describing polarizations, we follow the convention where the wave is observed from the point of view of the source. The biphoton state output from the cold atoms can be described as
\begin{eqnarray}
|\Psi_{1,2}\rangle=|HV\rangle\otimes(|\omega_s+\delta\rangle_1|\omega_{as}\rangle_2+|\omega_{as}+\delta\rangle_1|\omega_{s}\rangle_2). \nonumber \\
 \label{eq:BiphotonState12}
\end{eqnarray}
In our setup there is no length difference between path 1 and path 2 before the BS. The BS has the transformation $\hat{a}_3=(\hat{a}_{1H}-i\hat{a}_{2V})/\sqrt{2}$, $\hat{a}_4=(i\hat{a}_{1H}+\hat{a}_{2V})/\sqrt{2}$. The two-photon state with the Stokes photon in output port 3 and the anti-Stokes photon in output port 4 is
\begin{eqnarray}
|\Psi_{s3, as4}\rangle=\frac{1}{2}(|H, \omega_s+\delta\rangle_3|V, \omega_{as}\rangle_4 \nonumber \\
+|V, \omega_s\rangle_3|H, \omega_{as}+\delta\rangle_4),
 \label{eq:BiphotonState34}
\end{eqnarray}
which is a polarization-frequency-coupled entangled state \cite{CShu_arxiv2014}. Assuming the polarizers P$_3$ and P$_4$ in Fig.1 have the operations $\hat{T}_3=|H\rangle\langle P_3|$ and $\hat{T}_4=|H\rangle\langle P_4|$, respectively, we can decouple the polarization degree of freedom by the polarization projection:
\begin{eqnarray}
|\Psi_{s5, as6}\rangle=\hat{T}_3\hat{T}_4|\Psi_{s3, as4}\rangle\nonumber \\
=\frac{1}{2}|HH\rangle\otimes\big(\langle P_3|H\rangle\langle P_4|V\rangle|\omega_s+\delta\rangle_3|\omega_{as}\rangle_4 \nonumber \\
+\langle P_3|V\rangle\langle P_4|H\rangle|\omega_s\rangle_3 |\omega_{as}+\delta\rangle_4\big).
 \label{eq:BiphotonState341}
\end{eqnarray}
Setting $|$P$_3\rangle$=$\frac{1}{\sqrt{2}}$($|$H$\rangle$+$e^{-i\theta}$$|$V$\rangle$) and $|$P$_4\rangle$=$\frac{1}{\sqrt{2}}$($|$H$\rangle$+$|$V$\rangle$)=$|\nearrow\rangle$, we obtain the (unnormalized) two-photon state
\begin{eqnarray}
|\Psi_{s5, as6}\rangle=\frac{1}{4}|HH\rangle\otimes\big(|\omega_s+\delta, \omega_{as}\rangle+e^{i\theta}|\omega_s, \omega_{as}+\delta\rangle\big). \nonumber \\
 \label{eq:BiphotonState342}
\end{eqnarray}
This is a frequency-bin entangled two-photon state whose relative phase difference $\theta$ is transferred from the phase difference between the polarization bases H and V of $|$P$_3\rangle$.

Now we turn to realizing the polarizer operations $\hat{T}_3$ and $\hat{T}_4$. It is known that in the polarization space, any two pure polarization states can be connected by the combination of a quarter-wave plate (QWP) and HWP. The polarizer operation $\hat{T}_4=|H\rangle\langle \nearrow|$ can be realized by a HWP (with its fast axis aligned at 22.5-degree to H axis) and a PBS. For the complex polarizer operation $\hat{T}_3=\frac{1}{\sqrt{2}}|H\rangle(\langle H|+e^{i\theta}\langle V|)$, we use the combination of a QWP, a HWP and a PBS to project the desired polarization $|$P$_3\rangle$=$\frac{1}{\sqrt{2}}$($|$H$\rangle$+$e^{-i\theta}$$|$V$\rangle$) to the $|H\rangle$ linearly polarized output of the PBS.

\begin{figure}
\includegraphics[width=8.5cm]{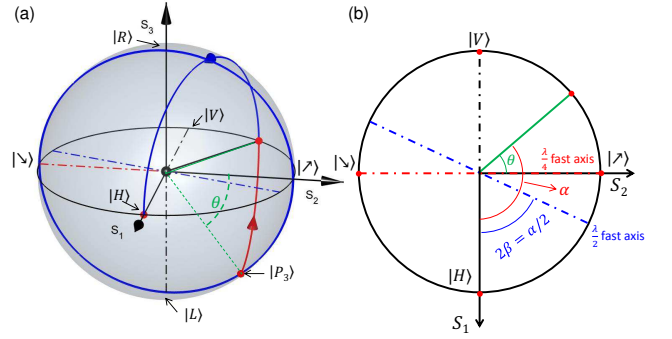}
\caption{\label{fig:Poincare Sphere}(Color online). (a) The Poincar$\acute{e}$ sphere. An arbitrary projection state can be rotated to $|H\rangle$ with the combination of a half-wave plate and a quarter-wave plate. Blue (red) dash-dot line represents rotation axis for half-wave plate (quarter-wave plate), and blue (red) arrow curve shows half-wave plate (quarter-wave plate) rotation. (b) The equatorial $S_{1}S_{2}$-plane view.}
\end{figure}

To illustrate this clearly, we plot the Poincar$\acute{e}$ sphere in Fig.~\ref{fig:Poincare Sphere}.
Any point on the Poincar$\acute{e}$ sphere represents a pure polarization state. Specifically, all linear polarization states are on the equator, with $|H\rangle$ and $|V\rangle$ at the intersections with $S_1$ axis, $|\nearrow\rangle=\frac{1}{\sqrt{2}}(|H\rangle+|V\rangle)$ and $|\searrow\rangle=\frac{1}{\sqrt{2}}(|H\rangle-|V\rangle)$ at the intersections with $S_2$ axis. The North Pole and South Pole represent the right-circular polarization $|R\rangle=\frac{1}{\sqrt{2}}(|H\rangle+i|V\rangle)$ and the left-circular polarization $|L\rangle=\frac{1}{\sqrt{2}}(|H\rangle-i|V\rangle)$, respectively. The projection state $|$P$_3\rangle$=$\frac{1}{\sqrt{2}}$($|$H$\rangle$+$e^{-i\theta}$$|$V$\rangle$) locates on the blue circle in the $S_{2}S_{3}$ plane. For a wave plate, its fast axis always locates on the equatorial plane and through the sphere centre, and the magnitude of rotation is determined by the phase retardance: 180$^{\circ}$ for a HWP and 90$^{\circ}$ for a QWP. We take the following two-step rotations to transfer the polarization state from $|$P$_3\rangle$=$\frac{1}{\sqrt{2}}$($|$H$\rangle$+$e^{-i\theta}$$|$V$\rangle$) to $|$H$\rangle$. First, we rotate $|$P$_3\rangle$ along the S$_2$-axis (fast axis of the QWP) with 90$^{\circ}$ and it becomes a linear-polarization state in the S$_1$S$_2$ plane with an angle $\alpha=\theta+\pi/2$ to the S$_1$ axis. Second, we set the fast axis of the HWP with an angle of $\alpha/2$ to the S$_1$ axis (i.e., we set the fast axis of the HWP with an angle of $\alpha/4=\theta/4+\pi/8$ to the H axis in the real space) and the intermediate polarization state is rotated to S$_1$ ($|H\rangle$). Therefore, to obtain the operation $\hat{T}_3=\frac{1}{\sqrt{2}}|H\rangle(\langle H|+e^{i\theta}\langle V|)$, we set the fast axes of the QWP and HWP with $\pi/4$ and $\beta=\alpha/4=\theta/4+\pi/8$, respectively, to the horizontal (H) axis.

We can also verify the above transformations by obtaining Jones matrix in the $\{|H\rangle,|V\rangle\}$ polarization bases:
\begin{eqnarray}
\hat{T}_3(\beta)=|H\rangle\langle P_3|=\frac{1}{\sqrt{2}}\left[
                                                         \begin{array}{cc}
                                                           1 & e^{i(4\beta-\pi/2)} \\
                                                           0 & 0 \\
                                                         \end{array}
                                                       \right],
 \label{eq:T3}
\end{eqnarray}
and
\begin{eqnarray}
\hat{T}_4=|H\rangle\langle \nearrow|=\frac{1}{\sqrt{2}}\left[
                                                         \begin{array}{cc}
                                                           1 & 1 \\
                                                           0 & 0 \\
                                                         \end{array}
                                                       \right],
 \label{eq:T4}
\end{eqnarray}

To determine the phase $\theta=4\beta-\pi/2$, we directly measure the two-photon beating in the time-domain. The two-photon temporal wave function can be obtained from Eq.(3):
\begin{eqnarray}
\Psi_{56}(t_5,t_6)=\langle HH;t_5,t_6|\Psi_{s5, as6}\rangle \nonumber \\
=\frac{1}{4}\psi_0(\tau)e^{-i\omega_s t_5}e^{-i\omega_{as}t_6}e^{-i\delta t_5}\big[1+e^{-i(\delta \tau-\theta)}\big],
 \label{eq:BiphotonState342}
\end{eqnarray}
where $\tau=t_6-t_5$ and $\psi_0(\tau)e^{-i\omega_s t_5}e^{-i\omega_{as}t_6}=\langle t_5, t_6|\omega_s, \omega_{as}\rangle$ is the Stokes-anti-Stokes biphoton temporal wave function generated from the SFWM process \cite{Du_JOSAB2008}. The coincidence measurement of two-photon is related to the Glauber correlation function
\begin{eqnarray}
G_{56}^{(2)}(\tau>0)=|\Psi_{56}(t_5,t_6)|^2 \nonumber \\ =\frac{1}{8}G_0^{(2)}(\tau)[1+\cos(\delta\tau-\theta)],
\label{eq:Glauber56}
\end{eqnarray}
where $G_0^{(2)}(\tau)=|\psi_0(\tau)|^2$ is the Glauber correlation function that can be measured before the BS. In the above discussion, we consider only Stokes photons in port 5 and anti-Stokes photons in port 6. Because of the system symmetry, Stokes photons can also appear in port 6 and anti-Stokes in port 5. Taking this into account, Eq. (\ref{eq:Glauber56}) is extended to $\tau<0$ and becomes
\begin{eqnarray}
G_{56}^{(2)}(\tau)=\frac{1}{8}G_0^{(2)}(|\tau|)[1+\cos(\delta|\tau|-\theta)].
\label{eq:Glauber56_2}
\end{eqnarray}
\begin{figure}
\includegraphics[width=8.5cm]{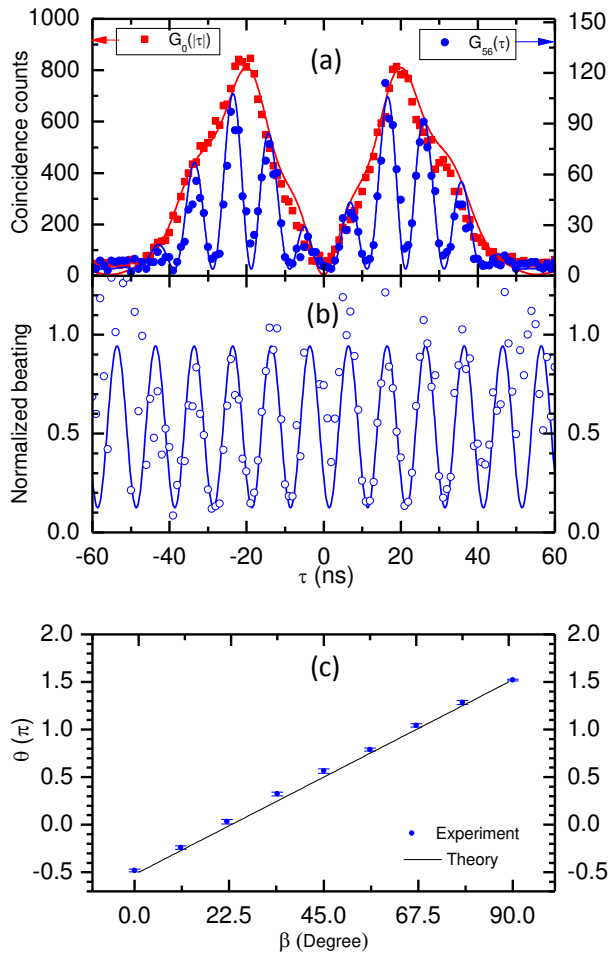}
\caption{\label{fig:RelativePhase}(Color online). (a) The biphoton waveform $G_0(|\tau|)$ at ports 1 and 2 before BS and the two-photon beating pattern $G_{56}(\tau)$ measured at ports 5 and 6. We set $\beta=79^{\circ}$, which corresponds to $\theta=1.25\pi$. (b) Normalized beating signal (blue circle) and sinusoidal fitting result (blue solid curve). The relative phase obtained from the fitted curve is $\theta_{exp}=(1.28\pm0.02)\pi$. (c) The relative phase $\theta$ of the frequency-bin entangled state as a function of the angle, $\beta$, between the fast axis of the HWP in $\hat{T}_3$ and the H axis. The phase values of the blue circles are obtained from sinusoidal fitting to the two-photon beatings. The black solid line is theoretically plotted from $\theta=4\beta-\pi/2$.}
\end{figure}

In Fig.~\ref{fig:RelativePhase}(a), the red square data show the measured biphoton temporal envelope $G_0(|\tau|)$ before the BS, which agrees well with the theoretical result (red solid curve) following the perturbation theory \cite{Du_JOSAB2008}. The blue dots show the two-photon beating pattern of a frequency-bin entangled two-photon state measured at port 5 and port 6, with $\beta=79^{\circ}$. The normalized experimental beating pattern is shown in Fig.~\ref{fig:RelativePhase}(b), from which we determine the relative phase $\theta_{exp}=(1.28\pm0.02)\pi$ by best fitting the data to a sine wave, which is consistent with $\theta=4\beta-\pi/2$. We obtain a visibility of $(77\pm6)\%$, which is beyond the requirement for violating the Bell inequality \cite{CHSH_PRL1969}. Figure \ref{fig:RelativePhase}(c) shows the relative phase $\theta$ as a function of $\beta$, the angle of the fast axis of the HWP in $\hat{T}_3$. The phase values (circle points) obtained from the two-photon beating agree well with the theoretical prediction (solid line).

\begin{figure}[h]
\includegraphics[width=8.5cm]{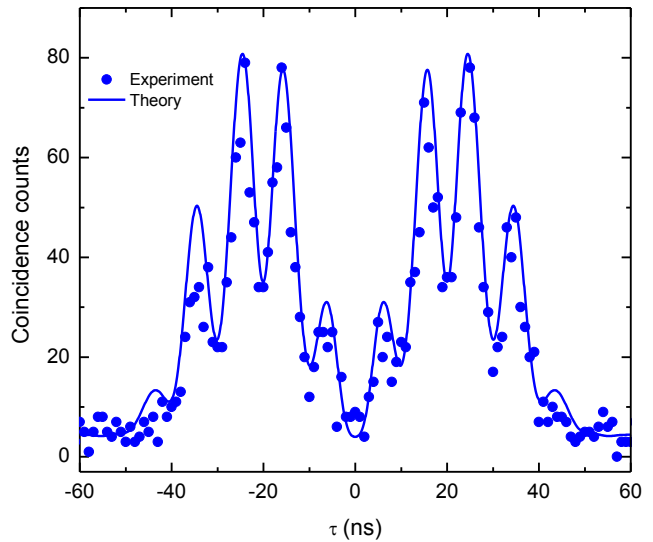}
\caption{\label{fig:ReducedVisibility}(Color online). The two-photon beating with $\phi=\pi/12$ and $\theta=\pi$. The experimental visibility is $(51\pm5)\%$. }
\end{figure}

To further show that this robust scheme also allows us to adjust the relative amplitude of the frequency-bin entanglement, we take $|$P$_3\rangle$=$\frac{1}{\sqrt{2}}$($|$H$\rangle$+$e^{-i\theta}$$|$V$\rangle$) and $|P_{4}\rangle=\cos\phi|H\rangle+\sin\phi |V\rangle $. The non-maximally entangled two-photon state and Glauber correlation function at the output ports 5 and 6 become
\begin{eqnarray}
|\Psi_{s5, as6}\rangle=\frac{1}{2\sqrt{2}}|HH\rangle\otimes\big(\cos\phi|\omega_s+\delta, \omega_{as}\rangle \nonumber \\
+\sin\phi e^{i\theta}|\omega_s, \omega_{as}+\delta\rangle\big),
 \\
G_{56}^{(2)}(\tau)=\frac{1}{8}G_0^{(2)}(|\tau|)\big[1+\sin2\phi\cos(\delta|\tau|-\theta)\big].
 \label{eq:BiphotonState56_2}
\end{eqnarray}
The visibility of the normalized beating is reduced to $V=|sin(2\phi)|$ due to the unbalanced amplitudes of the two bases. The experimental result under the condition of $\phi=\pi/12$ and $\theta=\pi$ is displayed in Fig.~\ref{fig:ReducedVisibility}, with a measured visibility of $(51\pm5)\%$.

In conclusion, we have demonstrated a polarization-frequency-coupled scheme for generating controllable frequency-bin entangled narrow-band biphotons, whose relative phase and amplitude between the entangled frequency bases can be arbitrarily tuned by adjusting the wave plates. Compared with the former schemes for wideband SPDC biphotons \cite{Bernhard_PRA2013, Zeilinger_PRL2009, Olislager_PRA2010}, our method only requires manipulating wave plates, which is more stable and versatile. With our recently developed temporal quantum-state tomography \cite{DuPRL2015}, we measured our biphoton bandwidth to be about 20 MHz. Our technique can be extended to generate subnatural-linewidth \cite{Du_PRL2008, LZhao_Optica2014, YanPRL2014} biphotons and find applications in realizing efficient light-matter quantum interface in a quantum network \cite{Kimble_Nature2008}.


The work was supported by the Hong Kong Research Grants Council (Project No. 16301214). C. S. acknowledges support from the Undergraduate Research Opportunities Program at the Hong Kong University of Science and Technology.


\end{document}